# Ferroelectric Switching in $ZnO/Zn_{1-x}Mg_xO$ Heterostructures: Atomistic Insights into Interfacial Coupling and Layer Architecture

Alireza Sepehrinezhad[1], Ali Mohammadi Dinani[2], Ece Gunay[3], Elizabeth C. Dickey[3], Jon-Paul Maria[4], Susan Trolier-McKinstry[4], Adri C.T. van Duin[4,2,1,*]

[1] Department of Mechanical Engineering, The Pennsylvania State University, University Park Pennsylvania 16802, United States

[2] Department of Chemical Engineering, The Pennsylvania State University, University Park, PA 16802, United States

[3] Department of Materials Science and Engineering, Carnegie Mellon University, Pittsburgh, Pennsylvania 15213, United States

[4] Department of Materials Science and Engineering, The Pennsylvania State University, University Park, PA 16802, United States

[*] Corresponding author email: acv13@psu.edu

## Abstract

Ferroelectric switching in heterostructures couples composition, layer topology, temperature, and interfacial boundary conditions. ReaxFF molecular dynamics isolates these variables in $ZnO/ Zn_{1-x}Mg_xO/ZnO$ and $Zn_{1-x}Mg_xO/ZnO/ Zn_{1-x}Mg_xO$ stacks. Within pristine, initially single-domain models, coupling to switchable $Zn_{1-x}Mg_xO$ reduces the applied field required to reverse ZnO by up to fivefold. Temperature generally lowers the coercive field, whereas Mg concentration produces a nonmonotonic response. At equal ZnO and $Zn_{1-x}Mg_xO$ (ZMO) proportions, structures with ZnO at the center switch at lower fields than those with ZMO at the center at all four temperatures examined, demonstrating a topology-dependent response. Layer-resolved trajectories reveal topology-dependent switching sequences with direction-dependent redistribution of normal stress near the heterointerfaces, consistent with a stress-assisted cooperative pathway. Limiting

MgO-containing structures exhibit sequential multilevel switching or low-polarity trapping, depending on thickness and temperature. Fixed-charge atomistic simulations complement previous continuum descriptions by resolving structural, energetic, and local stress evolution under a common applied field. TEM and STEM-EDS observations provide experimental structural context for the modeled architectures. Together, the results establish layer topology and interfacial mechanical confinement as design variables for wurtzite ferroelectric heterostructures.

## Introduction

Ferroelectric heterostructures provide a means of controlling polarization through the interfaces between dissimilar materials. Changes in composition, lattice structure, polarization, and electrostatic boundary conditions at an interface can modify the response of the complete stack and produce behavior that is absent from either constituent alone [1, 2]. Interface-mediated phenomena reported in oxide heterostructures include tunneling electroresistance, polarization-controlled electronic and magnetic states, and two-dimensional conducting layers [3–14]. These observations establish the interface as an active component of the material response rather than a passive boundary between layers.

A heterointerface can also be treated as a controlled perturbation to an otherwise continuous polar crystal. Its composition, position, layer sequence, and adjacent thicknesses can be prescribed independently, allowing the effects of chemical substitution and mechanical boundary conditions to be separated. This controllability is especially valuable for studying polarization switching because the measured response of an experimental film generally combines the effects of composition, interfaces, defects, electrodes, grain structure, local field distributions, and pre-existing domains. A controlled atomistic model can hold these factors fixed and vary one selected feature, thereby identifying how that feature changes the intrinsic response of the modeled system.

This approach is particularly relevant to wurtzite ferroelectrics, whose polarization reversal differs from that of perovskite ferroelectrics. Polar ZnO has a large switching barrier in its undoped state, whereas substitutional alloying can introduce local structural and strain fluctuations that lower the barrier and enable experimentally accessible switching [15-20]. In $Zn_{1-x}Mg_xO$ (ZMO), combined experimental and computational studies have linked ferroelectricity to the local strain fluctuations produced by Mg substitution [16, 17]. Composition and mechanical environment are therefore coupled at the atomic scale, making their separate contributions difficult to determine from an experimental hysteresis loop alone.

A direct experimental demonstration of interface-enabled switching was provided by Skidmore et al. [2]. Their study showed that polar but ordinarily non-switchable AlN and ZnO layers can reverse polarization when placed next to switchable $Al_{1-x}B_xN$, $Al_{1-x}Sc_xN$, or $Zn_{1-x}Mg_xO$ layers in bilayer and trilayer heterostructures. The proposed proximity-switching mechanism begins with nucleation in the ferroelectric layer, followed by propagation toward and across the internal interface. Elastic and electric fields associated with the advancing domain wall reduce the switching barrier in the adjacent non-ferroelectric layer. These experiments establish the phenomenon across several wurtzite material combinations, while also motivating atomically resolved studies of how composition, layer sequence, and interfacial mechanics influence the collective switching pathway.

Eliseev et al. subsequently developed a Landau–Ginzburg–Devonshire framework for proximity ferroelectricity in multilayer systems [21]. Their model describes each layer through a polarization-dependent free energy and solves the coupled polarization and electrostatic response self-consistently across the stack. The analysis shows that internal fields arising from polarization mismatch and depolarization can reshape the effective energy landscape of the constituent layers, producing either collective proximity switching or collective suppression depending on their properties and relative thicknesses. Applications of the theory to $Al_{1-x}Sc_xN$/AlN and $Zn_{1-x}Mg_xO$/ZnO bilayers demonstrate how electrostatic coupling can lower the effective switching barrier. This continuum description establishes the thermodynamic and self-consistent electrostatic basis for proximity switching and identifies general switching regimes. The present atomistic simulations complement that framework by resolving the layer-by-layer structural pathway and the accompanying evolution of local strain and stress, under an externally applied field; the fixed-charge treatment retains atomistic Coulomb interactions but does not resolve dynamic charge compensation or self-consistent layer-resolved field partitioning.

Phase-field modeling provides a complementary mesoscale description of domain nucleation, growth, and switching under coupled electrostatic and elastic boundary conditions [22-25]. It can access spatial and temporal scales beyond atomistic simulations and has been used to examine substrate constraint, dielectric heterogeneity, defects, interfaces, and strain-dependent domain evolution. Because polarization is represented as a continuous order parameter, phase-field models resolve collective domain behavior but do not directly describe discrete atomic rearrangements at a ZnO/$Zn_{1-x}Mg_xO$ interface. Thermodynamic, phase-field, and atomistic approaches therefore address different parts of the proximity-switching problem rather than serving as interchangeable descriptions.

Reactive molecular dynamics adds two capabilities needed for the present analysis. First, it resolves the atomic displacements, structural transformations, polarization evolution, and local stress distributions that accompany switching. Second, it permits controlled numerical experiments in which composition, topology, temperature, and initial structure can be varied independently. The ReaxFF potential used here was parameterized against density-functional-theory data for $Zn_{1-x}Mg_xO$ and reproduces the composition- and temperature-dependent switching trends in compositionally uniform ZMO structures established in earlier work [17]. Within this framework, polarization reversal emerges from the evolving interatomic forces and structure rather than from an imposed polarization free-energy function. The resulting trajectories can therefore reveal when and where different layers transform and how the local mechanical response develops during that sequence.

Here, we use ReaxFF molecular dynamics to perform a set of controlled comparisons across ZnO/ZMO/ZnO and ZMO/ZnO/ZMO heterostructures under a spatially uniform externally applied field, without self-consistent redistribution of that field in response to layer-dependent polarization and permittivity.

To determine composition dependence, Mg content is varied while topology and temperature are held fixed. To determine temperature dependence, temperature is varied for a fixed composition and topology. To determine topology dependence, the equal-proportion top-ZnO and mid-ZnO structures are compared at the same ZMO-layer composition and temperature. The latter comparison changes the layer sequence and its associated interfacial mechanical boundary condition while preserving the other principal simulation variables. The deliberately small, initially single-domain cells suppress microstructural and lateral domain-wall complexity, exposing the intrinsic vertically coupled response of each modeled stack. This controlled isolation of variables—not a reconstruction of every process present in an experimental film—is the central purpose of the atomistic analysis.

Across more than 400 simulations, this design reveals systematic composition, topology, and temperature dependences that would be difficult to separate experimentally. Increasing temperature generally lowers the simulated coercive field, whereas Mg content produces a non-monotonic response with a minimum near 40% Mg for several matched configurations. For the equal-proportion comparison, mid-ZnO stacks switch at lower applied fields than their top-ZnO counterparts at each temperature examined, showing that layer sequence affects the collective response when composition is held fixed. Several heterostructures also switch at lower externally applied fields than either corresponding standalone constituent within the same simulation framework. The switching trajectories show that these reductions coincide with layer-resolved transformation sequences and localization of normal stress near the heterointerfaces. We therefore identify a stress-assisted cooperative switching pathway in which interfacial mechanical interactions contribute to the response without asserting that stress is the only possible contribution.

The simulations further identify multistep polarization responses in selected 60%-Mg and MgO-containing model structures. Rather than treating the fully Mg-substituted cases as practical target compositions, we use them as limiting systems that probe how the MgO compositional endpoint changes the collective response. Depending on its relative thickness and temperature, this layer produces either sequential switching with intermediate polarization plateaus or suppression of sustained ferroelectric switching. These limiting cases map the transition between cooperative switching, multistep response, and proximity suppression within a common atomistic framework. The fixed-charge treatment used in this work does not describe dynamic charge compensation or self-consistent macroscopic field partitioning; accordingly, the analysis is restricted to the directly resolved structural, polarization, energetic, and stress evolution under the same applied-field protocol.

# Methods

Two symmetric heterostructure topologies were constructed, as shown in Figure 1. In the top-ZnO configuration, a central $Zn_{1-x}Mg_xO$ layer is enclosed by two ZnO layers. In the mid-ZnO configuration, a central ZnO layer is enclosed by two $Zn_{1-x}Mg_xO$ layers of equal composition and thickness. The outer-to-middle thickness ratio ($t_o/t_m$), was varied from 0.2 to 5 in both topologies. These paired architectures enable the influence of layer sequence and its associated mechanical boundary conditions to be examined while composition and temperature are held fixed.

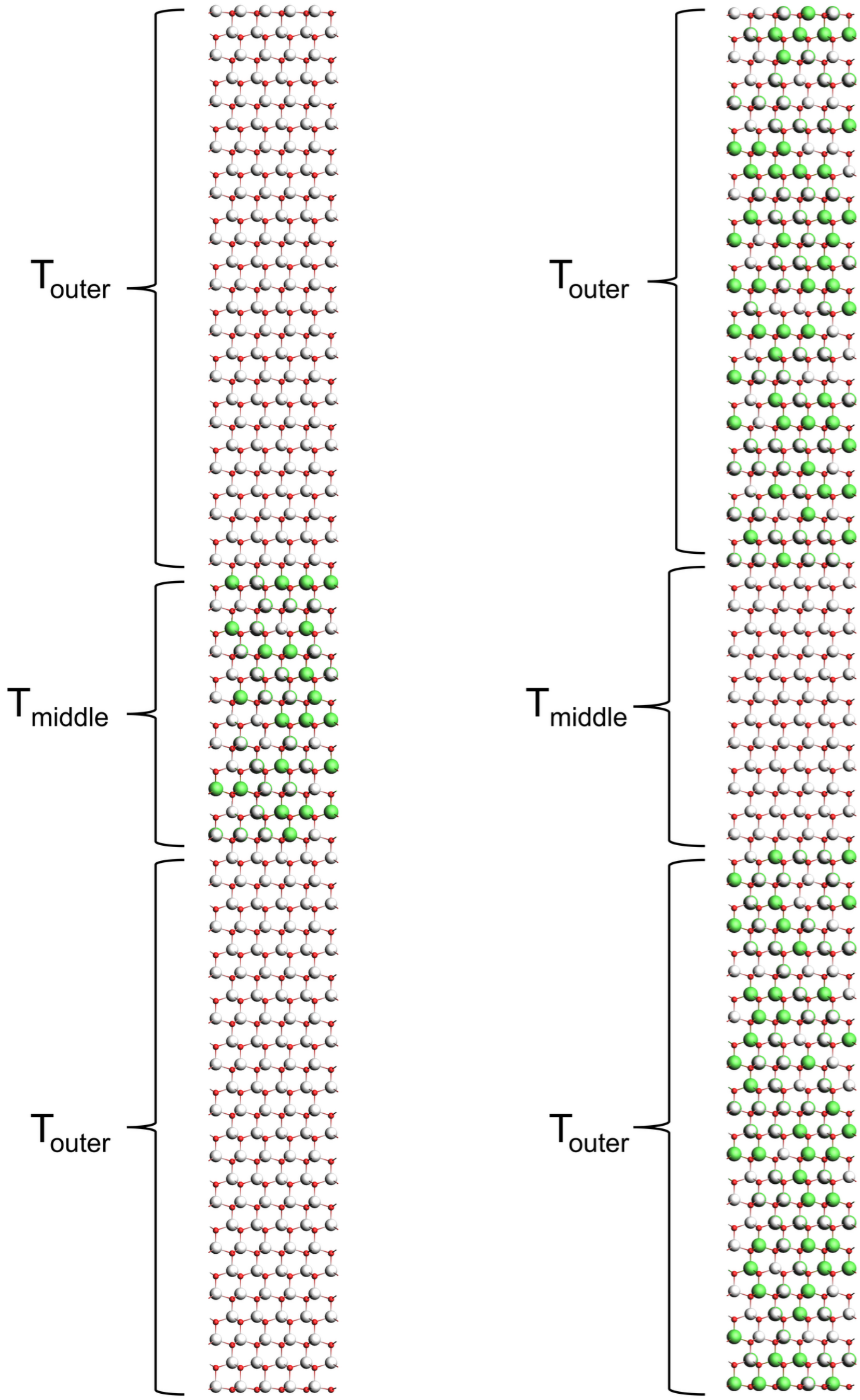

**Figure 1: Atomistic models of the top-ZnO and mid-ZnO heterostructures. The top-ZnO configuration (left) contains a central $Zn_{0.6}Mg_{0.4}O$ layer between two ZnO layers.** The mid-ZnO configuration (right) contains a central ZnO layer between two $Zn_{0.6}Mg_{0.4}O$ layers. $T_{\text{outer}}$ and $T_{\text{middle}}$ denote the thicknesses used to define the ratio $t_o/t_m$. Mg and Zn atoms are shown in green and silver, respectively, and their displayed radii are enlarged for clarity.

The simulation cells contained ~2000 atoms and had lateral dimensions of 11.46 Å and 13.24 Å, in X- and Y-directions, respectively. Periodic boundary conditions were applied along X- and Y-directions, with free surfaces and vacuum layers along the polar direction. The polar axis was oriented along Z-direction.

The systematic composition series included Mg molar fractions of $x = 0, 0.2, 0.4$, and 0.6. Selected $x = 1$ structures were additionally simulated as limiting MgO-containing cases to determine how the fully Mg-substituted endpoint changes the collective response. Composition dependence was evaluated by varying x at fixed topology, thickness ratio, and temperature. Temperature dependence was evaluated at fixed composition and topology, while topology dependence was evaluated using the equal-proportion top-ZnO and mid-ZnO pair at the same ZMO-layer composition and temperature. This design allows each principal variable to be examined while the others are held constant.

A sinusoidal electric field was applied uniformly along the polar axis to obtain the polarization response of each structure. As established in the previous $Zn_{1-x}Mg_xO$ ReaxFF study [17], conventional Electronegativity Equalization Method (EEM) charge equilibration under an applied field produced unphysical long-range charge transfer between the free surfaces and an artificial opposing field. To prevent this artifact, the structures were first equilibrated and the resulting species-resolved average partial charges were then fixed during electric-field cycling. This protocol retains field–charge coupling, atomistic Coulomb interactions, and the field-driven structural response while applying the same electrical driving condition to every configuration. Polarization was calculated using the procedure described in Ref. [17], based on the modern theory of polarization [26].

The ReaxFF parameterization developed and validated in Ref. [17] was used for all simulations. That work benchmarked the force field against density-functional-theory calculations across the $Zn_{1-x}Mg_xO$ compositional range and identified structural signatures of the intermediate nonpolar configuration, including tetrahedral flattening and bond-angle symmetrization. The fixed species-resolved charges used during field cycling were approximately +1.054 for Mg and +0.974 for Zn. Oxygen atoms bonded to at least one Mg atom had a slightly larger charge magnitude than the remaining oxygen atoms, with representative values of −1.0001 and −0.9821, respectively. Because these charges remain fixed during cycling, the simulations do not model dynamic charge compensation or self-consistent macroscopic field partitioning according to the permittivities of the individual layers. A uniform externally applied field is therefore the common control variable used to compare the simulated structures; it is not reconstructed as the layer-resolved field in an experimental capacitor.

The resulting database comprises more than 400 molecular dynamics simulations spanning the two layer topologies, multiple thickness ratios and Mg compositions, and temperatures of 100, 300, 500, and 700 K. All simulations were performed in the isothermal–isobaric ensemble with a time step of 0.25 fs. Temperature and pressure were controlled using Berendsen thermostat and barostat coupling constants of 100 and 2500 fs, respectively. Each simulation was maintained at its specified temperature throughout equilibration and electric-field cycling.

Instantaneous per-atom stress tensors were calculated using the approach developed and reported in [27]. For a system with N particles and volume V, the global stress tensor P is defined as:

$$\boldsymbol{PV} = \langle \sum_{\boldsymbol{i=1}}^{\boldsymbol{N}} \boldsymbol{m_i v_i \otimes v_i} + \boldsymbol{W}(\boldsymbol{r^N}) \rangle \quad \textbf{(1)}$$

where $\boldsymbol{m_i}$ and $\boldsymbol{v_i}$ are the mass and instantaneous velocity of the i$^{th}$ particle. The symbol $\otimes$ is for the dyadic tensor product. W is the global virial tensor, and can be obtained through different approaches, discussed in detail in [27].

Cross-sectional TEM samples were prepared by wedge polishing using an Allied Multiprep II system, followed by final thinning by Ar+ ion milling in a Gatan PIPS II with minimal exposure to reduce beam-induced damage. S/TEM imaging was performed on an aberration-corrected Thermo Scientific Themis microscope operated at 200 kV. Selected area electron diffraction (SAED) patterns were acquired from the ZnO and ZMO layers to verify the crystallographic phase and orientation. Annular dark-field (ADF) images were collected using a detector with a collection angle range of 46-200 mrad. EDS elemental mapping was performed using a Super-X G2 detector with an estimated probe current of~110 pA.

The simulation cells were intentionally initialized as defect-free, single-domain structures with small lateral dimensions. This design suppresses competing contributions from heterogeneous nucleation, pre-existing domain walls, extended defects, and complex lateral domain-wall motion. Consequently, the calculated coercive fields represent switching within this constrained atomistic system and are not expected to reproduce experimental coercive fields quantitatively. The value of the model lies in the controlled comparisons: every paired structure is evaluated with the same cell construction, charge treatment, applied-field protocol, and analysis procedure. Differences produced by changing composition, temperature, thickness ratio, or layer topology can therefore be assigned to that controlled change within the model. In particular, the equal-proportion top-ZnO and mid-ZnO comparison isolates the effect of layer sequence and the resulting interfacial mechanical environment on polarization switching and stress evolution.

### *Statistical Analysis*

No inferential statistical analysis was performed. Each reported coercive field and time-resolved stress trajectory was obtained from a single molecular dynamics trajectory for the specified composition, topology, thickness ratio, and temperature; values are therefore not presented as ensemble means, and no error bars are assigned. Comparative trends were evaluated across simulations conducted using the same cell-construction, charge-treatment, and field-cycling protocols.

In this work, ReaxKit (version 1.0.0) [28] was used to organize and post-process the simulation outputs, while plots were generated using Matplotlib, and Microsoft Excel was used for data tabulation and final quality-control checks. ReaxKit is a modular Python toolkit for preparing, parsing, organizing, and analyzing ReaxFF molecular dynamics simulations . It converts engine-specific simulation outputs into consistent, analysis-ready data structures and supports reproducible workflows encompassing data processing, aggregation, visualization, and reporting.

## Results and Discussion

### *Effects of composition, temperature, and layer topology on heterostructure switching*

Mg substitution in ZnO introduces local strain fluctuations associated with cation-size mismatch, thereby facilitating polarization reversal in ZMO [16]. Within the same defect-free simulation framework, compositionally uniform ZnO requires an applied field of approximately 15 $MV.cm^{-1}$ to reverse [17]. Incorporating ZnO into a ZnO/ZMO/ZnO heterostructure reduces the field required to reverse the ZnO layers by as much as a factor of five. The ZMO layer begins transforming first, after which the adjacent ZnO layers follow through an interface-initiated, domain-wall-mediated sequence. This controlled comparison demonstrates that the heterostructure environment fundamentally alters the switching response of ZnO; the accompanying evolution of interfacial stress is examined in the following subsection.

Figure 2 maps the coercive field $E_c$ of the top-ZnO configuration as a function of the outer-to-middle thickness ratio ($t_o/t_m$), temperature, and Mg concentration in the ZMO layer; corresponding results for the mid-ZnO configuration are presented in Figure S1. The simulation matrix separates the selected dependencies through controlled comparisons: temperature is varied at fixed composition and architecture, ZMO-layer composition is varied at fixed temperature and thickness ratio, and layer-fraction effects are evaluated at fixed ZMO composition and topology. Because the cell dimensions and simulation protocol are otherwise unchanged, these comparisons isolate the effects of the selected variables within the modeled single-domain system.

The coercive fields reported in Figure 2 are defined by the externally applied, spatially uniform field at which the net polarization of the simulated slab crosses zero; they should therefore not be interpreted as local electric fields within the individual layers. Although the heterostructure coercive field commonly lies between the corresponding values for compositionally uniform ZMO and ZnO, several configurations switch at an applied field below both standalone materials. Because the heterostructures and their constituent reference systems are evaluated using the same field protocol, this result directly demonstrates that the switching response of the coupled stack cannot be represented as a simple interpolation between its components. In the atomistic trajectories, the reduced switching field coincides with reversal of the ZMO layer, localization of stress near the heterointerfaces, and subsequent reversal of the ZnO layers, supporting a stress-assisted cooperative switching pathway.

The controlled trends in Figure 2 separate three contributions to the heterostructure response. First, decreasing $t_o/t_m$ increases the relative fraction of the more readily switchable ZMO layer and generally lowers the coercive field. Second, $E_c$ decreases with increasing temperature, consistent with thermally assisted polarization reversal [17]. Third, the dependence on Mg concentration is nonmonotonic: among the $x = 0.20$, 0.40, and 0.60 compositions examined, $x=0.40$ generally produces the lowest coercive fields over the sampled thickness ratios and temperatures. The separation among the three compositions becomes small under several conditions, including $t_o/t_m=0.33$ and 1 at 500 K, indicating that thermal activation and layer fraction can reduce the sensitivity to ZMO composition. Together, these trends show that the heterostructure coercive field is jointly governed by composition, relative layer thickness, and temperature rather than by the overall Mg content alone.

The layer-resolved trajectories reveal the structural sequence underlying the cooperative response. The Mg-rich middle layer first evolves toward an intermediate nonpolar configuration, after which the adjacent ZnO-rich layers undergo the corresponding transition and complete the polarization reversal. The atomistic simulations therefore resolve the order and spatial progression of the structural transformations. Because the fixed-charge treatment does not include dynamic charge redistribution, it does not determine how bound charge associated with a polarization discontinuity would be screened or how long the intermediate configuration would remain stabilized in an experimental heterostructure. The conclusions drawn here concern the directly observed structural pathway rather than the electrostatic lifetime of its intermediate states.

The present results share two broad trends with the continuum analysis of Eliseev et al. [21]. Both studies predict a proximity response in which a switchable ZMO layer promotes polarization reversal in adjacent ZnO, and both show an overall decrease in coercive field as the relative fraction of ZMO increases. The correspondence should nevertheless be interpreted at the level of these qualitative trends rather than as evidence of an identical microscopic mechanism. The continuum model resolves self-consistent electrostatic and phenomenological coupling, whereas the present fixed-charge MD simulations resolve atomic rearrangement, local strain, and stress evolution under a common applied-field protocol. Moreover, the trilayer simulations exhibit nonmonotonic thickness dependence and, in selected configurations, coercive fields below that of standalone ZMO. These departures from the simplest bilayer trend motivate the explicit examination of layer topology and interfacial mechanical coupling below.

The $x=1$ systems containing pure MgO layers are included as limiting compositional tests of the heterostructure response rather than as representations of readily accessible experimental architectures. These endpoint systems determine whether the trends established for ZMO persist as the Mg content approaches its compositional limit. They exhibit two distinct responses—retention near an intermediate nonpolar configuration and multilevel, nonbinary hysteresis—which are analyzed separately in the later discussion of MgO-rich heterostructures.

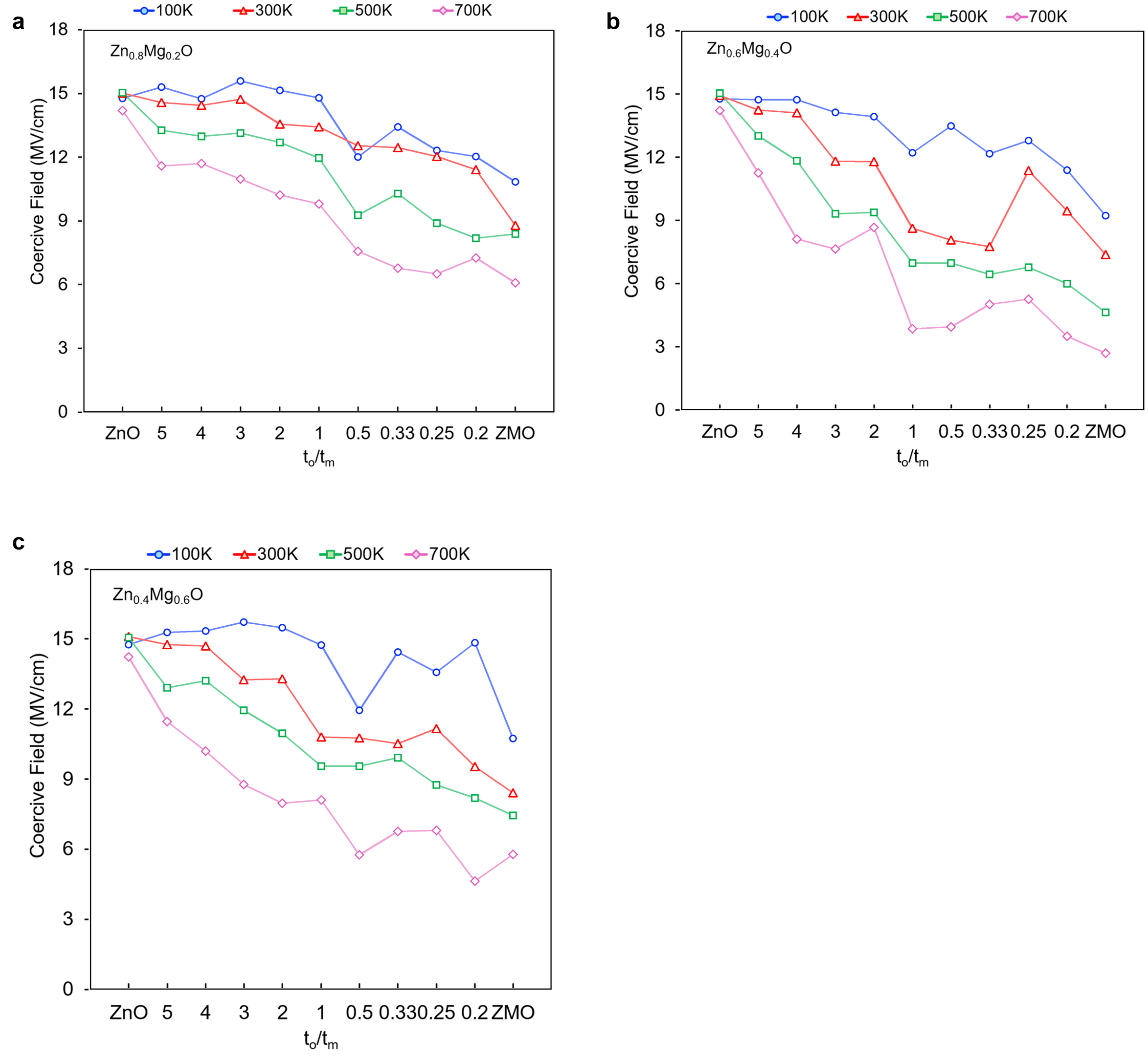

**Figure 2: Coercive field ($E_c$) of the top-ZnO heterostructures as a function of the outer-to-middle thickness ratio ($t_o/t_m$) and temperature for ZMO middle layers containing (a) 20%, (b) 40%, and (c) 60% Mg.** The cell dimensions, layer topology, and simulation protocol are held constant within each controlled comparison. Solid lines connect the simulated values at each temperature as guides to the eye.

Figure 3 compares top-ZnO and mid-ZnO heterostructures containing equal proportions of ZnO and ZMO. The two structures have the same layer dimensions and ZMO composition but differ in the placement of the ZnO and ZMO layers. At every temperature examined, the coercive-field ratio exceeds unity, indicating that the mid-ZnO structure switches at a lower field than the corresponding top-ZnO structure.

The topology dependence is consistent with the different mechanical boundary conditions imposed on ZnO in the two architectures. In the mid-ZnO configuration, the central ZnO layer contacts ZMO on both surfaces and is therefore confined between two heterointerfaces. In the top-ZnO configuration, each outer ZnO layer has one ZnO–ZMO interface and one free surface, providing an additional pathway for structural relaxation. The central ZnO layer consequently sustains greater in-plane tensile strain in the mid-ZnO architecture, whereas the outer ZnO layers can partially relax in the top-ZnO architecture. When composition and temperature are held fixed, this difference in confinement is associated with the lower coercive field of the mid-ZnO structures. The stress distributions examined in the following subsection provide the spatially resolved test of this mechanical interpretation.

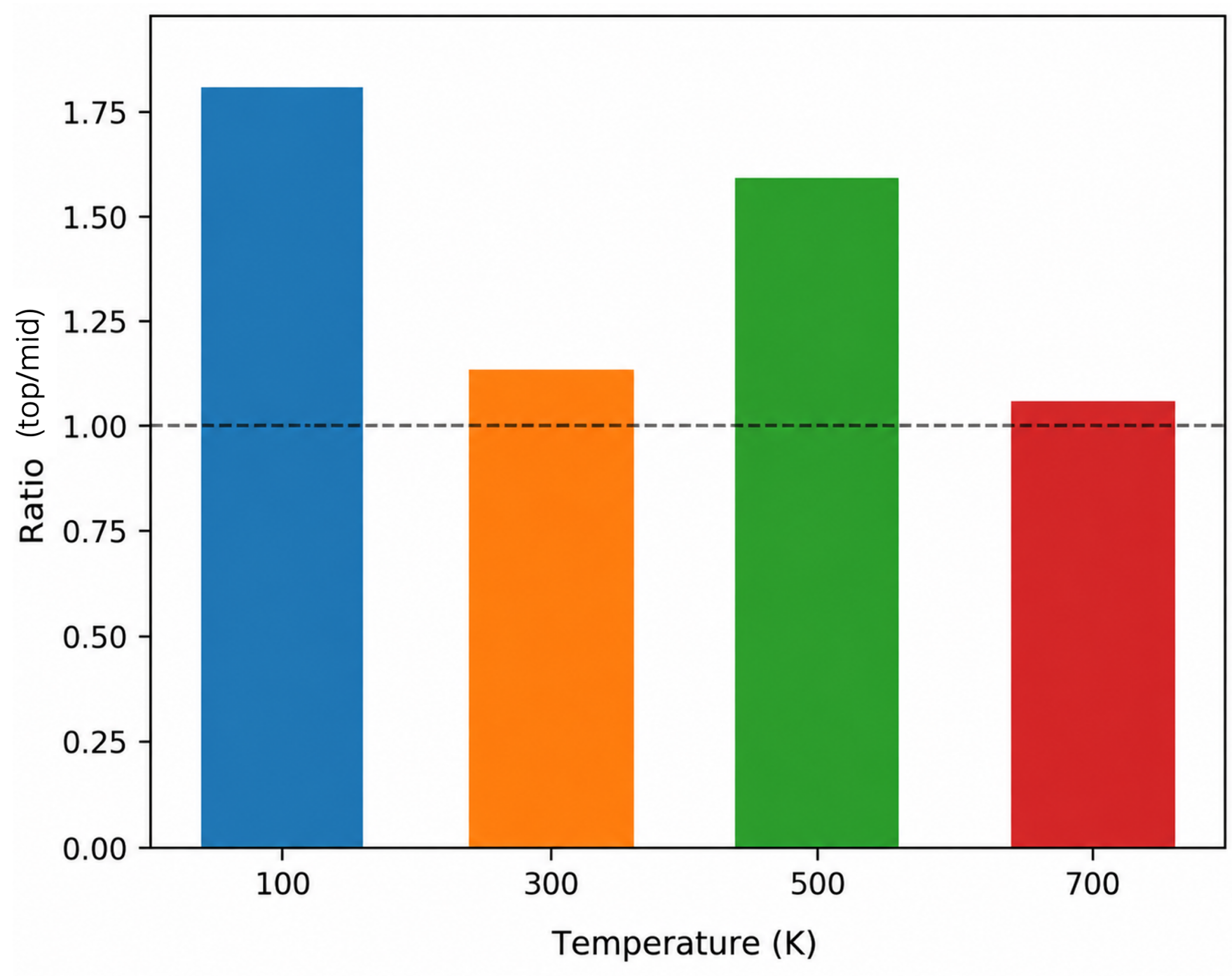

**Figure 3: Ratio of the coercive fields of top-ZnO and mid-ZnO heterostructures at $t_o/t_m = 0.5$, containing equal proportions of ZnO and ZMO, $E_c^{\text{top-ZnO}}/E_c^{\text{mid-ZnO}}$.** The structures have the same layer dimensions and ZMO composition but differ in the placement of the ZnO and ZMO layers. Ratios greater than unity indicate a lower coercive field for the mid-ZnO architecture. Each value is obtained from one MD trajectory rather than an average over independent statistical replicas.

*Interfacial stress localization during switching initiation and propagation*

The atomistic cells contain coherent ZnO/ZMO interfaces, allowing mechanical perturbations generated near an interface to be transmitted through the continuous layered structure. The experimental cross-sectional TEM images show columnar grains extending through the ZnO and ZMO layers, while the STEM-EDS maps show an abrupt compositional transition and a uniform Mg distribution within the ZMO layer. These observations support the layered geometry and compositional architecture represented in the atomistic models.

To resolve the coupling between electrical switching and local mechanical response, Figure 4 follows a representative top-ZnO heterostructure containing 40% Mg in the central ZMO layer. The structure was simulated at 700 K with $t_o/t_m$=0.2 and has a calculated coercive field of 3.50 MV.cm$^{-1}$. Figure 4a shows the applied sinusoidal field together with the potential energy of the slab. The abrupt energy changes near the coercive fields mark field-driven structural relaxation during polarization reversal. The inset shows the corresponding symmetric P–E loop, with a remanent polarization of approximately 90 μC.cm$^{-2}$.

Figure 4b compares the evolution of the region-averaged local ($\sigma_{zz}$) stress in the bulk and at the two heterointerfaces. Pronounced interfacial stress excursions occur near the coercive fields, marked by stars. During switching in the (+Z) direction, the bottom-interface stress changes earlier and reaches a larger excursion than the top-interface stress. The asymmetry reverses during switching in the (-Z) direction, when the top-interface response becomes dominant. This reversal of the stress asymmetry follows the observed direction of the polarization-reversal front: upward under a positive field and downward under a negative field. The plotted values are local virial stresses normalized by atomic volumes and therefore describe the spatial and temporal distribution of atomistic stress rather than a macroscopic applied stress. No dislocation nucleation or glide was observed during the analyzed trajectory.

As the system approaches each coercive field, the top- and bottom-interface stress responses diverge, producing a direction-dependent mechanical asymmetry across the middle ZMO layer. Comparison of Figures 4a and 4b shows that this redistribution of interfacial stress coincides with the abrupt energetic changes and zero-polarization crossings associated with switching. The controlled trajectory therefore shows that interfacial stress redistribution and polarization reversal are dynamically coupled. The observed correspondence supports a mechanical contribution to interface-proximate switching initiation, although the trajectory alone does not separate that contribution quantitatively from the other interactions operating within the heterostructure.

The spatial correspondence is shown in Figure 4c, which maps the local $\sigma_{zz}$ stress and polarization orientation in the XZ plane at the positive and negative coercive fields, respectively. The white dashed lines identify the heterointerfaces. At $+E_c$, the larger stress excursion near the bottom interface coincides with the first locally reversed polarization vectors, after which the reversal front propagates upward. At $-E_c$, the corresponding sequence begins near the top interface and propagates downward. In both directions, the first reversed region appears at or near the more strongly perturbed interface, while the remaining layers reverse sequentially. The switching process is therefore spatially heterogeneous even within the laterally compact cell, and its initiation and propagation direction are correlated with the evolving interfacial stress distribution. Recent large-scale simulations of ZnO/ZnMgO architectures likewise identify buried interfaces as preferential filament-nucleation regions, although those calculations emphasize the reconstructed switching-front electrostatics rather than the local stress evolution examined here [29].

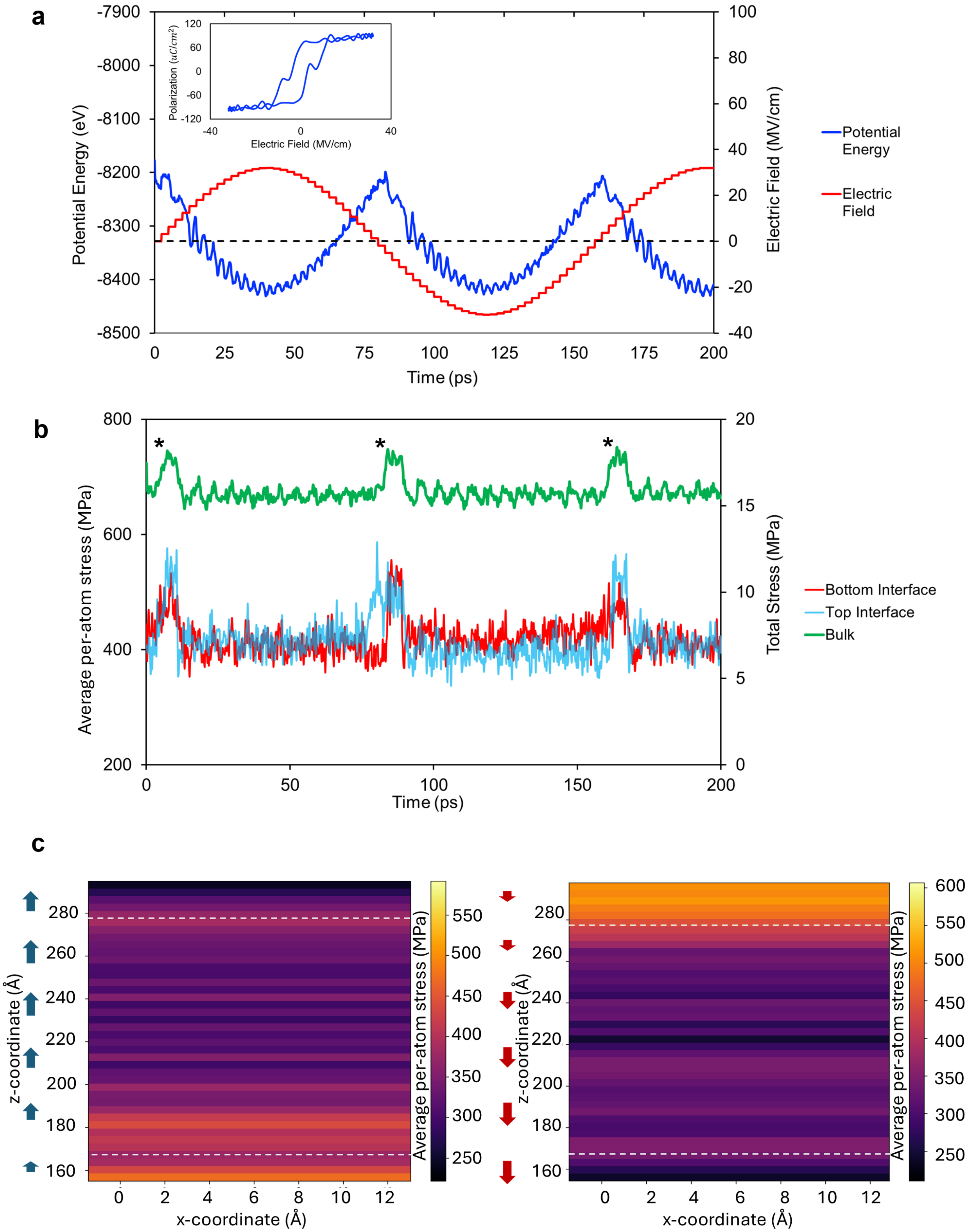

**Figure 4. Coupled electrical, energetic, and local mechanical response of a representative $ZnO/Zn_{0.6}Mg_{0.4}O/ZnO$ top-ZnO heterostructure at 700 K with $t_o/t_m$=0.2.** (a) Potential energy and applied electric field during the 200 ps trajectory. The inset shows the resulting P–E hysteresis loop. (b) Region-averaged local ($\sigma_{zz}$)

stress in the bulk and at the bottom and top heterointerfaces. Stars mark the times at which the net polarization crosses zero. (c) Spatial distributions of the local ($\sigma_{zz}$) stress in the XZ plane at 5 ps, corresponding to the first positive coercive field, and 54.25 ps, corresponding to the negative coercive field. Arrows indicate the local polarization orientation, and white dashed lines mark the heterointerfaces. The local stress values are obtained from the atomic virial normalized by the assigned atomic volume and should not be interpreted as macroscopic applied stresses.

To determine whether the switching-associated response is strongest along the polar axis, Figure 5 compares the three normal stress components at the bottom interface. The two in-plane components, $\sigma_{xx}$ and $\sigma_{yy}$, exhibit similar temporal behavior, whereas the out-of-plane component, $\sigma_{zz}$, undergoes the largest fluctuations and switching-associated excursions. The interfacial stress response is therefore anisotropic, with its strongest variation occurring along the polarization direction. Rapid fluctuations are superimposed on this field-driven response because the local per-atom stress also contains contributions from thermal vibrations and short-timescale structural rearrangements. Consequently, the switching-associated response is identified from the coordinated stress excursions near the coercive fields rather than from individual instantaneous peaks.

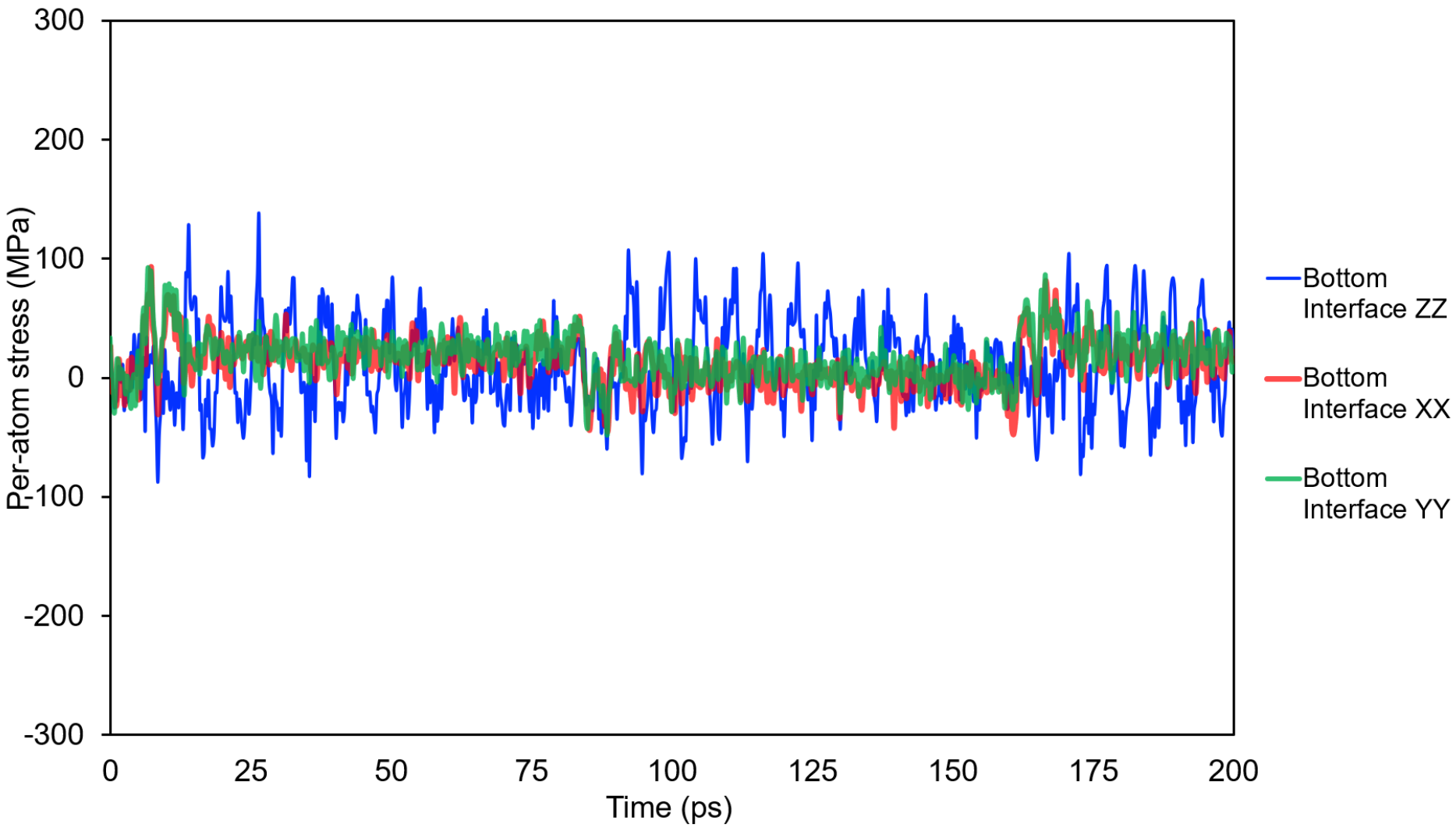

**Figure 5: Time evolution of the region-averaged local normal stress components $\sigma_{xx}$, $\sigma_{yy}$, and $\sigma_{zz}$ at the bottom interface of the ZnO/$Zn_{0.6}Mg_{0.4}O$/ZnO heterostructure shown in Figure 4.** The structure was simulated at 700 K with $t_o/t_m$=0.2.

Across the sampled compositions, temperatures, and thickness ratios, the first locally reversed region occurs either at a heterointerface or within a few unit cells of it. The robust result is therefore preferential interface-proximate switching initiation rather than a fixed atomic nucleation plane or a universal threshold value of local stress. The small variation in the initiation position reflects differences in local composition, thermal fluctuations, and layer geometry across the controlled simulation matrix. Quantifying a site-specific nucleation probability or critical local-stress distribution would require larger lateral cells and independent trajectory sampling; within the present systems, the recurring correspondence between interface-proximate stress excursions and switching initiation establishes the propagation trend.

### *Composition-induced sequential switching and multilevel hysteresis*

Nonbinary polarization responses emerge in two classes of the simulated structures: top-ZnO heterostructures containing a high-concentration-Mg layer and compositionally uniform ZMO films containing 60% Mg. Figure 6 examines the first class using the limiting ZnO/MgO/ZnO structure at 300 K and $t_o/t_m$=0.5. At the same topology, temperature, and thickness ratio used for the lower-Mg systems, replacing the central ZMO layer with MgO changes the response from a conventional binary reversal to a sequence containing multiple polarization plateaus. This controlled compositional endpoint therefore reveals a switching pathway in which different parts of the heterostructure transform at distinct times during each field cycle.

The intermediate polarization plateaus correspond directly to layer-resolved structural configurations sampled during the simulated switching trajectory. They occur when different regions of the heterostructure have reached different stages of reversal and therefore establish a sequential structural pathway. The fixed-charge treatment does not include the dynamic charge redistribution that could stabilize a polarization discontinuity in an experimental film. The plateaus should therefore be interpreted as transient structural states under the applied field, rather than as evidence of a resolved electronic screening mechanism. Large-scale reactive molecular dynamics simulations of pristine, Mg-modified, and layered ZnO show that reversal proceeds through spatially rugged inversion-boundary filaments that nucleate at free surfaces or buried ZnO/ZnMgO interfaces, advance predominantly along the field direction, and subsequently expand laterally and coalesce [29]. The approximately planar front observed here can therefore be interpreted as a laterally confined limit of this broader nucleation-and-growth morphology rather than as a universal domain-wall geometry.

Figure 6a resolves the origin of the multilevel response in the ZnO/MgO/ZnO heterostructure. Two distinct changes in potential energy, labeled 1 and 2, occur during each half-cycle of the applied field. These features coincide with separate structural transformations of the central MgO and outer ZnO regions, demonstrating that the constituent layers do not respond simultaneously. The preference of the MgO region for a more symmetric, weakly polar or nonpolar coordination makes its field-driven response distinct from that of the polar ZnO layers. Between the two transformations, the slab occupies a mixed layer-resolved configuration in which one region has transformed while the other has not, producing the intermediate plateau in the P–E loop. Repetition of this sequence during field reversal generates the observed multistep hysteresis response.

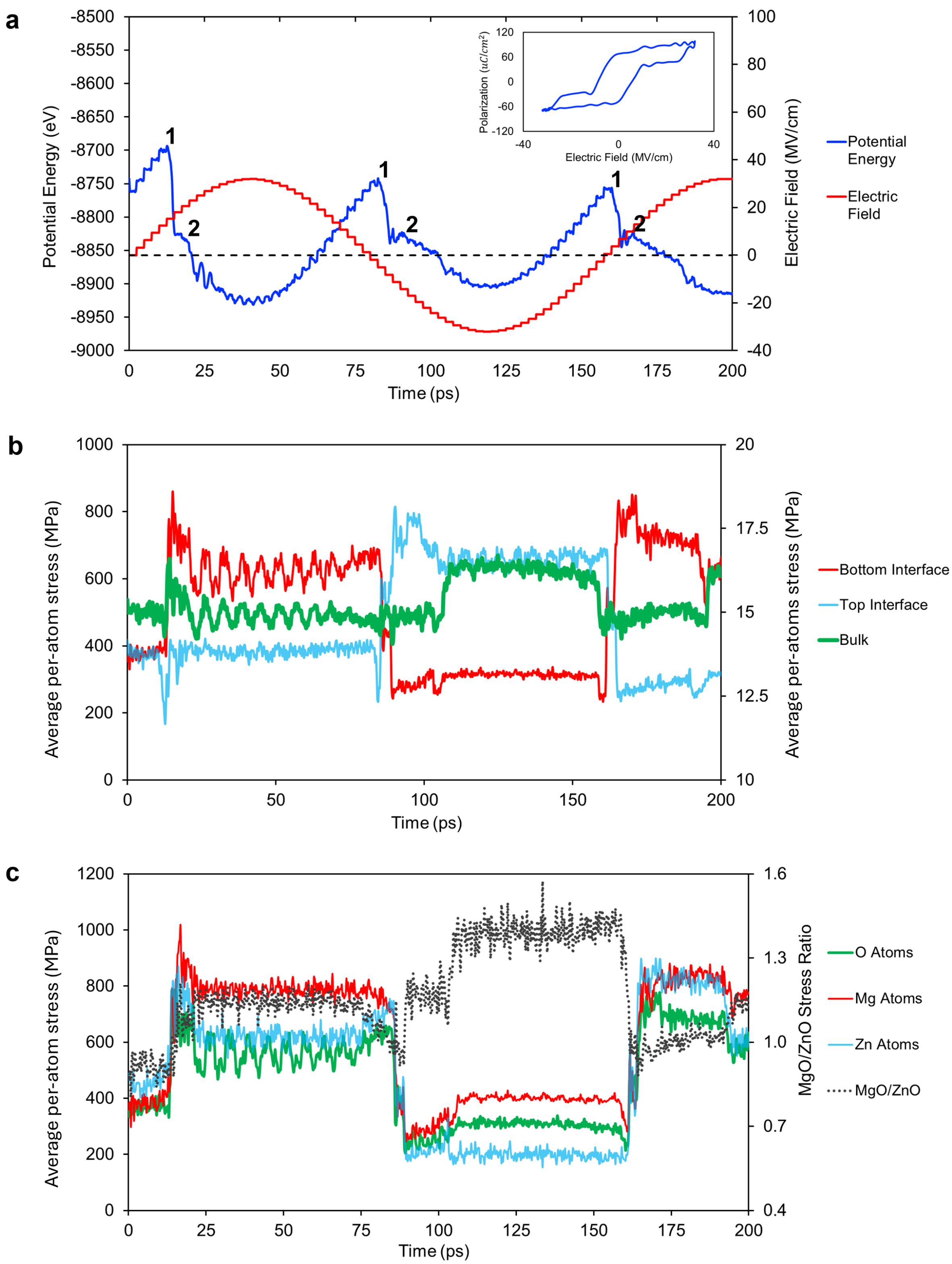

**Figure 6: Sequential switching and multilevel hysteresis in a ZnO/MgO/ZnO top-ZnO heterostructure simulated at 300 K with $t_o/t_m$=0.5.** (a) Potential energy and applied electric field during the 200 ps trajectory. Labels 1 and 2 mark the two successive structural transformations observed during each half-cycle. The inset shows the

resulting P–E loop with four polarization levels sampled during field cycling. (b) Region-averaged local ($\sigma_{zz}$) stress in the bulk and at the bottom and top heterointerfaces. (c) Species-resolved local ($\sigma_{zz}$) stress for O, Mg, and Zn atoms at the bottom interface. The dotted curve shows the ratio of the mean Mg-atom stress to the mean Zn-atom stress.

The multistep structural response is accompanied by a corresponding redistribution of local stress. In Figure 6b, the separation between the top- and bottom-interface stress traces changes abruptly at the times of the two structural events identified in Figure 6a and persists while the system occupies the intermediate polarization plateau. The species-resolved analysis in Figure 6c further shows that the Mg and Zn atoms carry different local stresses during these transformations, with the Mg atoms exhibiting the larger stress over substantial portions of the switching cycle. The abrupt changes in the Mg-to-Zn stress ratio demonstrate that the structural transitions redistribute mechanical loading between the chemical sublattices. Within this composition-controlled limiting case, the unequal stress partitioning provides an atomistic connection between the identity of the central layer and the emergence of the sequential electrical response.

### *Suppression of reversible switching in MgO-rich heterostructures*

The compositionally limiting heterostructures containing a central MgO layer exhibit a second response regime in which reversible polarization switching is suppressed. Figure 7 shows a representative ZnO/MgO/ZnO top-ZnO structure simulated at 100 K with $t_o/t_m$=0.2. Under these conditions, the central MgO layer constitutes the dominant fraction of the stack. After an initial field-driven transformation, the net polarization remains near zero over the subsequent field cycles, and the P–E response becomes nearly linear. The outer ZnO layers therefore do not retain their independently switchable response; instead, their behavior becomes coupled to the low-polarity configuration adopted by the MgO-rich core (Figure 7a).

Figure 7b identifies the structural event that produces this response. The potential energy and Z-direction displacement change abruptly during the first half-cycle, indicating a simultaneous energetic relaxation and contraction of the slab. Both quantities then remain near their new values despite continued cycling of the applied field. No subsequent polarization reversal is observed during the remainder of the 200 ps trajectory. The structure is therefore trapped, on the simulated timescale and within the applied-field range, in the weakly polar or nonpolar intermediate reached during the initial transformation.

The trapping behavior is consistent with the composition-dependent energy landscape established previously for $Zn_{1-x}Mg_xO$ [17]. Those equation-of-state calculations showed that Mg-rich compositions increasingly favor a more symmetric, low-polarity configuration over the polar wurtzite structure. In the present heterostructure, the central MgO region accesses this low-polarity coordination and constrains the response of the adjacent ZnO layers, which no longer display separate switching events after the initial transformation. The trajectory supports identification of the resulting structure as a weakly polar or nonpolar intermediate; a more specific crystallographic assignment requires a separate local-coordination analysis.

Comparisons at fixed layer topology show that the response of the MgO-containing structures depends jointly on temperature and the relative MgO thickness. As the central MgO fraction increases, the remanent polarization decreases and the response evolves from reversible binary switching to sequential multilevel switching and, ultimately, to trapping in the nonpolar intermediate. Within the configurations that continue to reverse, an apparent coercive field can still be defined from the zero-polarization crossing. Once cyclic reversal is lost, however, the coercive field no longer provides an adequate measure of the response, and the disappearance of remanence and reversibility becomes the defining result.

The progression between these regimes reflects competition between the structural preferences of the constituent layers. At intermediate MgO fractions, the ZnO and MgO regions can transform sequentially while still completing reversal during each field cycle, producing the multilevel response described above. When the MgO-rich region becomes sufficiently dominant, its preference for the nonpolar intermediate prevents the adjacent ZnO layers from restoring a collectively polar state. The stack then remains trapped in the low-polarity configuration for the duration of the trajectory rather than completing repeated switching cycles.

The simulations consequently identify a compositional and geometric window for multilevel switching: the Mg-rich region must be sufficiently influential to separate the layer transformations in time, but not so dominant that it traps the complete stack in the nonpolar intermediate. This balance provides a design principle that can be tested in experimentally accessible high-Mg ZMO heterostructures seeking multilevel polarization responses while retaining remanence and reversible field cycling. More generally, the results suggest that increasing the difference between the switching kinetics of adjacent layers may enlarge the temporal separation between their transformations and widen the intermediate polarization plateaus.

Figure 7c shows that the initial transformation is accompanied by a sharp divergence between the top- and bottom-interface stress responses. After the structure enters the nonpolar intermediate, the interfacial stresses continue to vary under the oscillating field, but these variations no longer produce polarization reversal. Interfacial stress localization is therefore not sufficient by itself to sustain switching; it must remain coupled to a structural state capable of completing reversible polar transformation. This contrast sharpens the interpretation of the earlier results: stress redistribution participates in the switching pathway, but the available composition-dependent structural states determine whether reversal can proceed.

Following the initial transformation, the bulk stress settles to a nearly constant value while the two interfaces continue to exhibit field-dependent variations. The local interfacial response is thus no longer transmitted through the slab as a collective structural reversal. Instead, the dominant MgO-rich region mechanically accommodates the low-polarity configuration, decoupling the oscillating interfacial response from the net polarization of the stack. This mechanical signature is consistent with the persistent energy, displacement, and polarization responses in Figure 7b.

Figure 7d shows that the average interfacial stress in the suppressed-switching case is lower than in the conventional- and multilevel-switching cases. This reduction is consistent with stress delocalization accompanying formation of the intermediate low-polarity configuration, which lacks the polar distortion associated with localized interfacial stress.

The defining feature of the trapped MgO-rich configurations is the absence of sustained cyclic switching within the simulated field range, rather than an elevated or reduced coercive field. After the initial transition, the applied field continues to modulate the interface stresses but no longer drives a corresponding bulk deformation or polarization reversal. The combined energy, displacement, polarization, and stress trajectories therefore show that the low-polarity MgO-rich configuration interrupts the cooperative switching pathway that enables reversal in the ZnO/ZMO heterostructures.

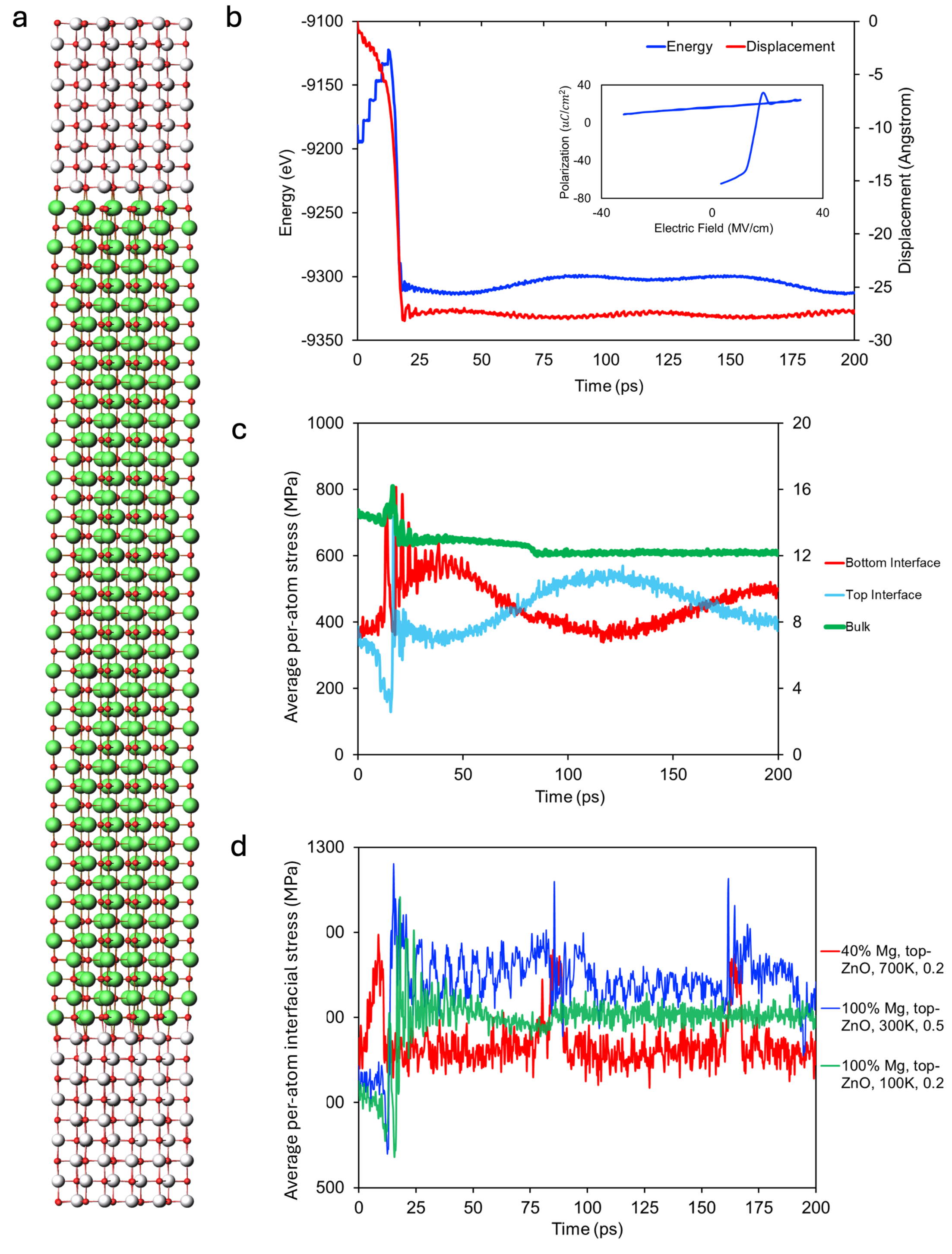

**Figure 7: Suppression of reversible switching in a ZnO/MgO/ZnO top-ZnO heterostructure simulated at 100 K with $t_o/t_m$=0.2.** (a) Atomic configuration at 92.5 ps after the initial field-driven transformation into the low-polarity intermediate. Zn, Mg, and O atoms are shown in white, green, and red, respectively. (b) Potential energy and Z-

direction displacement during the 200 ps trajectory. The inset shows the corresponding nearly linear P–E response after the initial transformation. (c) Region-averaged local ($\sigma_{zz}$) stress in the bulk and at the bottom and top heterointerfaces. The structure undergoes one initial transformation but does not recover cyclic polarization reversal during the remaining field cycles. (d) Comparison of the average interfacial per-atom stress for three representative response regimes: conventional switching (red), multilevel switching (blue), and suppressed reversible switching following the initial transformation (green).

### *Composition dependence of coercive field and interfacial stress*

To isolate the effect of ZMO-layer composition, we compare three top-ZnO heterostructures containing 20%, 40%, and 60% Mg in the middle layer while holding the temperature at 300 K and the thickness ratio at $t_o/t_m$=0.5. The layer topology, cell dimensions, and electric-field protocol are otherwise identical. The calculated coercive fields are 12.5, 8.05, and 10.75 MV.cm$^{-1}$ for the 20%, 40%, and 60% Mg structures, respectively. The dependence is therefore nonmonotonic: among the sampled compositions, the 40% Mg structure has a coercive field approximately 36% lower than that of the 20% Mg structure and 25% lower than that of the 60% Mg structure. The intermediate composition consequently provides the most readily switchable heterostructure in this controlled series, demonstrating that the response is governed by the composition-dependent local environment rather than by Mg concentration alone.

Figure 8 connects this nonmonotonic coercive-field dependence to the evolution of local stress. Because all three structures experience the same time-dependent electric field, the timing of their stress excursions can be compared directly. The switching-associated bottom-interface stress excursion occurs first in the 40% Mg structure, showing that it reaches the switching-associated interfacial stress state at a lower applied field than the 20% and 60% Mg structures. The correspondence between the earliest stress response and the lowest $E_c$ supports a composition-dependent coupling between the local strain environment and polarization reversal. Because composition is the only intentionally varied parameter in this series, the comparison controls for temperature, thickness ratio, and topology when evaluating this relationship.

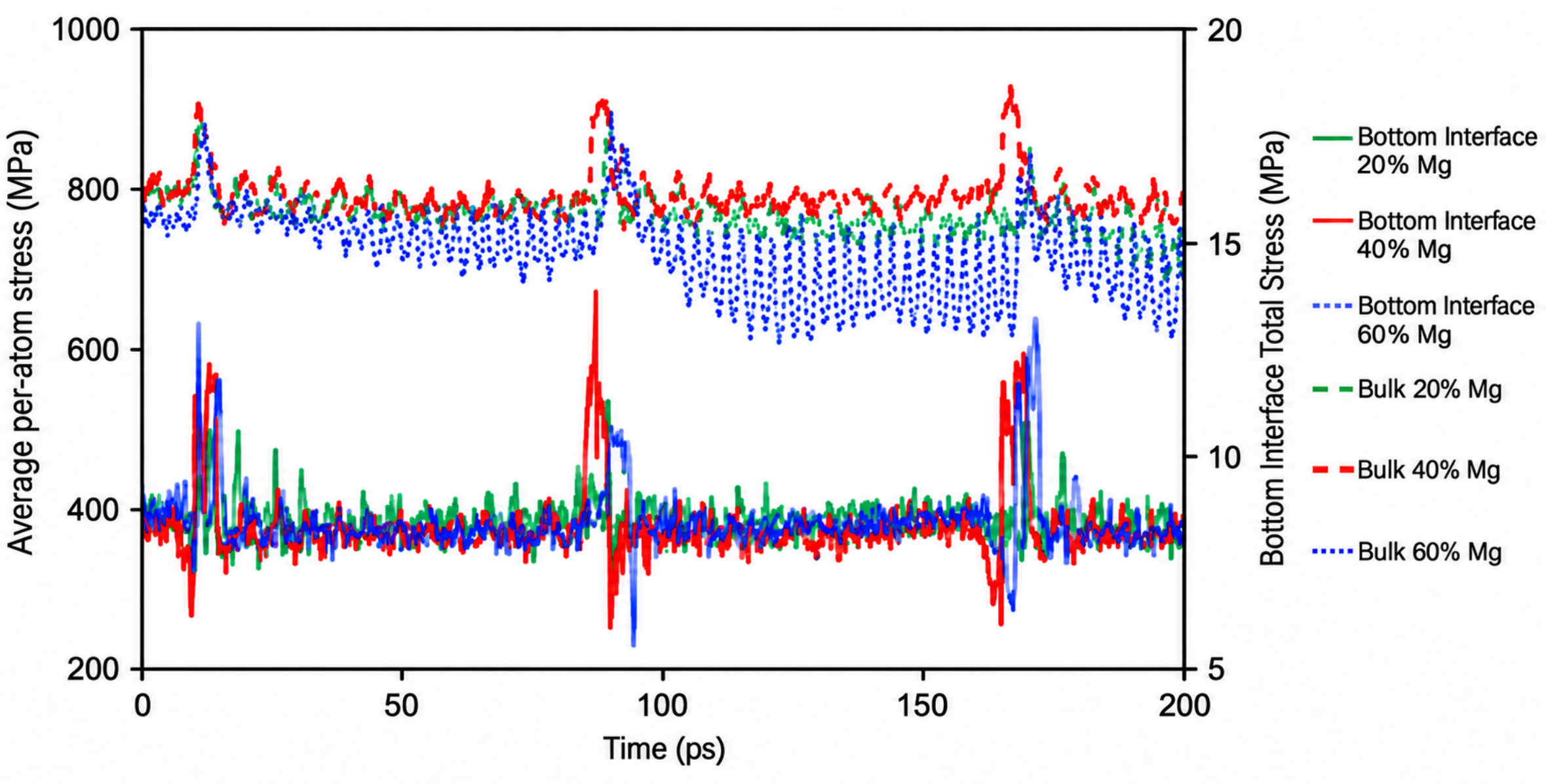

**Figure 8: Effect of ZMO-layer Mg concentration on the evolution of local $\sigma_{zz}$ stress in ZnO/$Zn_{1-x}Mg_xO$/ZnO top-ZnO heterostructures.** The 20%, 40%, and 60% Mg structures are simulated at 300 K with $t_o/t_m$=0.5 under the same time-dependent electric field. Solid curves show the region-averaged local stress at the bottom interface, and dashed curves show the corresponding bulk response. The switching-associated stress excursion occurs earliest for the 40% Mg structure, which also exhibits the lowest coercive field in the controlled series.

## *Experimental structure of the ZnO-ZMO interfaces*

Figure 9 provides the experimental structural context for the modeled ZnO/ZMO interfaces. The cross-sectional bright-field TEM image in Figure 9a shows a multilayer stack with columnar grains extending through the ZnO and $Zn_{1-x}Mg_xO$ layers. The selected-area electron-diffraction (SAED) patterns acquired from the ZnO and ZMO layers are consistent with the wurtzite structure, with the indexed out-of-plane (0002) reflections indicated in the inset. The ADF-STEM image and corresponding STEM-EDS elemental maps in Figure 9b further resolve the chemistry of the heterostructure. The ZnO and ZMO layers are distinguished by their elemental contrast, while Mg signal is confined to the ZMO layer. The transition between the ZnO and ZMO regions appears abrupt within the spatial resolution of the measurement, with Mg distributed uniformly throughout the ZMO layer. Together, these observations support the layered geometry, structural character, and compositional architecture represented in the atomistic models.

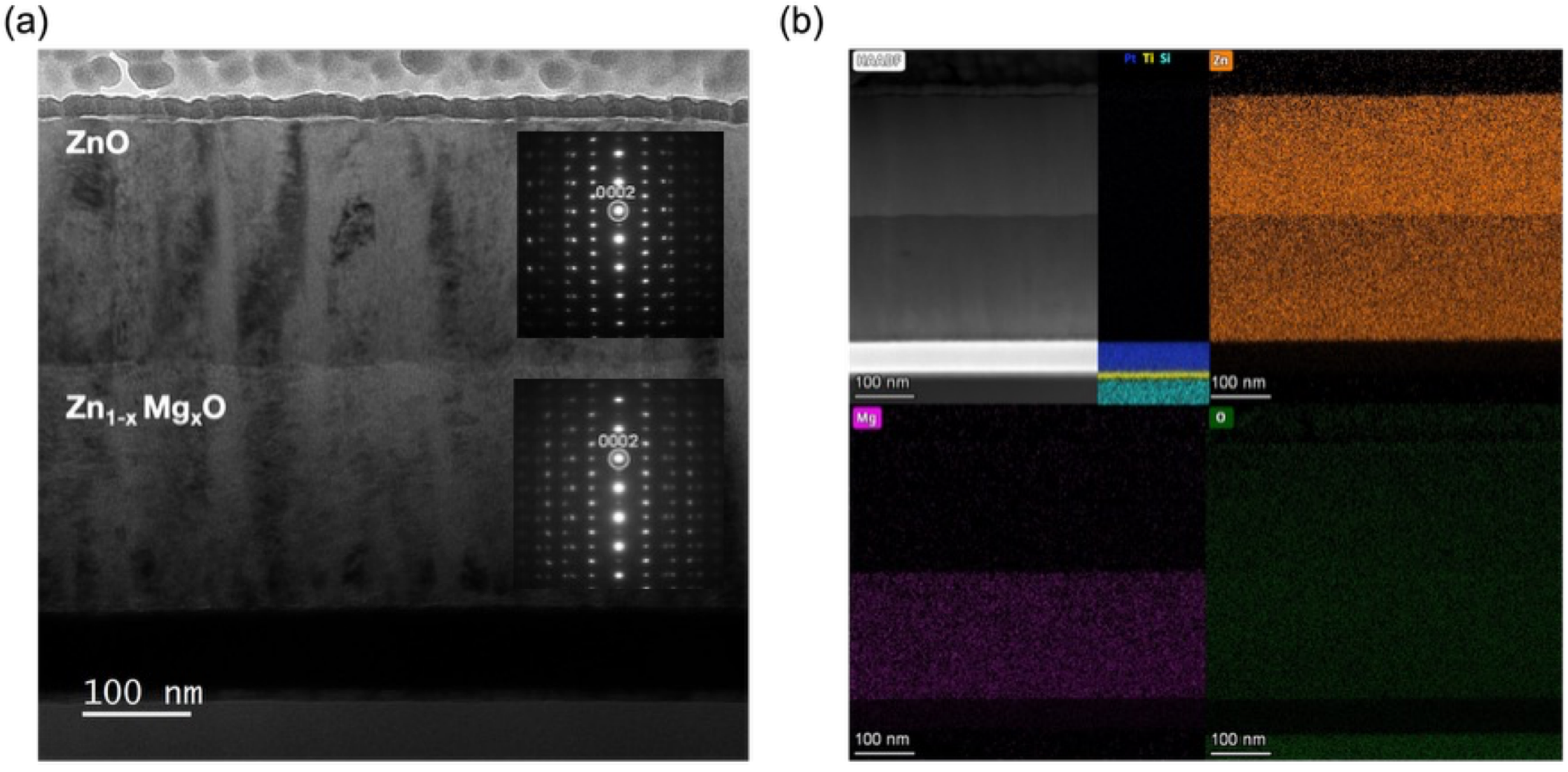

**Figure 9: Experimental structural characterization of a ZnO/ZMO heterostructure.** (a) Cross-sectional bright-field TEM image of a ZnO//ZMO//Pt(111)//Ti//$SiO_2$ stack. The insets show SAED patterns acquired from the ZnO and ZMO layers, both consistent with the wurtzite structure; the indexed (0002) reflections are marked. (b) ADF-STEM image from the same region with corresponding STEM-EDS elemental maps for Zn, Mg, O, Pt, Ti and Si. The maps distinguish the ZnO and ZMO layers and show an approximately uniform Mg distribution within the ZMO layer, together with a compositionally abrupt ZnO/ZMO interface.

The microscopy and atomistic simulations serve complementary roles in this study. The microscopy shows columnar grains extending through the ZnO and ZMO layers, an abrupt ZnO/ZMO compositional boundary, and a uniform Mg distribution within the ZMO layer. Using this structural context, the simulations independently vary composition, layer topology, temperature, and relative thickness to isolate how each variable modifies polarization reversal. The experimental observations therefore ground the geometry and compositional architecture of the models, while the predicted cooperative switching behavior and coercive-field trends provide specific targets for subsequent electrical characterization of ZnO/ZMO heterostructures.

## Conclusions

This study uses atomistic molecular dynamics as a controlled numerical experiment to separate the principal variables governing polarization reversal in $Zn_{1-x}Mg_xO$ heterostructures. Composition, relative layer thickness, temperature, and layer topology were varied independently while the remaining conditions were held fixed. The small, defect-free, initially single-domain systems thereby expose the intrinsic through-thickness coupling between the constituent layers and provide direct access to the sequence of structural transformation, polarization reversal, and local-stress redistribution.

Across more than 400 simulations, placing ZnO in a heterostructure with switchable $Zn_{1-x}Mg_xO$ reduced the applied field required to reverse ZnO by as much as a factor of five relative to uniform ZnO within the same simulation framework. In several configurations, the complete heterostructure switched at an externally applied field below the coercive fields of both corresponding standalone constituents, establishing a cooperative response that cannot be inferred from either constituent alone. Increasing temperature generally reduced the coercive field, whereas Mg concentration produced a nonmonotonic dependence, with $x = 0.40$ yielding the lowest coercive fields among the sampled compositions. In the equal-proportion comparison, mid-ZnO stacks switched at lower fields than their top-ZnO counterparts at all four temperatures examined. Because composition and temperature were matched within each pair, this result identifies layer topology—and the accompanying difference in mechanical confinement—as a contributor to the switching response. The layer-resolved trajectories provide a mechanistic interpretation of these trends. The $Zn_{1-x}Mg_xO$ region first evolves toward a non-polar intermediate configuration, after which reversal propagates into the adjacent ZnO. This sequence is accompanied by localized, direction-dependent excursions in normal stress at the heterointerfaces, supporting a stress-assisted cooperative switching pathway without assigning the response to stress alone. The limiting MgO-containing systems further map the boundaries of this behavior: depending on temperature and relative layer thickness, they exhibit either sequential layer transformations that produce multilevel hysteresis or trapping in a low-polarity intermediate that suppresses reversible switching over the simulated timescale.

The fixed-charge treatment retains field–charge coupling, Coulombic interactions, atomistic structural evolution, and local mechanical response under a common applied field. It does not describe dynamic charge compensation or self-consistent layer-resolved macroscopic field partitioning, and the absolute coercive fields remain sensitive to the finite dimensions, defect-free structures, and molecular-dynamics timescale. The comparative trends are nevertheless obtained consistently within the same framework. TEM and STEM-EDS observations provide complementary structural context by showing columnar grains extending through the ZnO and $Zn_{1-x}Mg_xO$ layers, an abrupt compositional transition at their boundary, and a uniform Mg distribution within the alloy layer. Taken together, the results show that a heterointerface is an active mechanical and structural component of the switching pathway: its composition and position can promote cooperative reversal, separate the transformations of neighboring layers, or stabilize a low-polarity state.

## Acknowledgments

The molecular dynamics simulations, film growth, and electron microscopy were supported by the Center for 3D Ferroelectric Microelectronics Manufacturing ($3DFeM^2$), an Energy Frontier Research Center funded by the U.S. Department of Energy, Office of Science, Office of Basic Energy Sciences, under Award No. DE-SC0021118. EG and ECD acknowledge the use of the Materials Characterization Facility at Carnegie Mellon University, MCF-677785.

## Conflict of Interest

The authors declare no conflict of interest.

## Data Availability Statement

The data supporting this study are available upon reasonable request through the Penn State Materials Computation Center's online request portal: https://www.mri.psu.edu/materials-computation-center/connect-mcc.

## Generative AI Use

OpenAI ChatGPT was used to assist with language editing, manuscript organization, and consistency checking. The authors reviewed and revised all generated text and take full responsibility for the scientific content and final manuscript.